\documentclass[12pt,A4paper]{article}

\usepackage{t1enc} % accent
\usepackage[francais]{babel} % français
\usepackage{amsmath}
\usepackage{epsfig,amssymb}
\usepackage{graphicx,psfrag}

\usepackage{color}

\begin{document}

\title{\bf Intermittency in aging}
\author{L. Buisson, L. Bellon, S. Ciliberto  \\
Ecole Normale Sup\'erieure de Lyon, Laboratoire de Physique,\\
C.N.R.S. UMR5672,  \\ 46, All\'ee d'Italie, 69364 Lyon Cedex
07,  France\\
        }
 \maketitle

\begin{abstract}
 The fluctuation-dissipation relation (FDR) is measured on the
dielectric properties of  a gel (Laponite)  and of a polymer glass
(polycarbonate). For the gel it is found that during the
transition from a fluid-like to a solid-like state the fluctuation
dissipation theorem is strongly violated. The amplitude and the
persistence time of this violation are decreasing functions of
frequency. Around $1Hz$ it may  persist for several hours. A very
similar behavior is observed in  polycarbonate after a quench
below the glass transition temperature. In both cases the origin
of this violation is a highly intermittent dynamics characterized
by large fluctuations. The relevance of these results for recent
models  of aging are discussed.
\end{abstract}

\medskip
{\bf PACS:} 75.10.Nr, 77.22Gm, 64.70Pf, 05.20$-$y.

\section{ Introduction}

Many systems in nature, such as glasses, spin-glasses, colloids
and granular materials, present an extremely slow relaxation
towards equilibrium and, when external conditions are modified,
the physical properties of these materials evolve as a function of
time: they are aging.  For example, when a glassy material is
quenched from above its glass transition temperature $T_g$ to a
temperature lower than $T_g$, any response function of the
material depends on the time $t_w$ elapsed from the quench
\cite{Struick}. Another example of aging  is given by
colloidal-glasses, whose properties evolve during the sol-gel
transition which may last several days \cite{Kroon}. An important
feature of  aging materials is the dependence of their physical
properties on the thermal history of the  sample. Indeed
experimental procedures, based on multiple cycles of cooling,
heating and waiting times, have shown the existence of  two
spectacular effects: memory and rejuvenation. Specifically, aging
materials  present a rejuvenation for any negative temperature
perturbation and at the same time during heating they remind the
stops at fixed temperature done during cooling (see for example
\cite{Jonason,bellonM} and references therein). Several models
have been proposed to explain such a beahvior but from an
experimental point of view it is not easy to distinguish between
them. The above mentioned experimental procedures have  been,
indeed, extremely useful to fix several constrains for the
phenomenological models\cite{trap,droplet}, but
 these procedures are mainly based on the study of the response of
the system to an external perturbation. Therefore they are unable
to give new insight  on the system  dynamics. Let us consider for
example the trap model\cite{trap} which is based on a phase space
description. Its basic ingredient is an activation process and
aging is associated to the fact that deeper and deeper valleys are
reached as the system evolves \cite{bertin}. The dynamics in this
model has to be intermittent because either nothing moves or there
is a jump between two traps. This contrasts, for example, with
mean field dynamics which is continuous in time\cite{Kurchan}.
Therefore, from an experimental point of view, it is extremely
important to study not only the response of the system but also
its thermal fluctuations.  This analysis is also related  to
another important aspect of aging dynamics, that is the definition
of the temperature. Indeed recent theories \cite{Kurchan} based on
the description of spin glasses by a mean field approach proposed
to extend the concept of temperature using a Fluctuation
Dissipation Relation (FDR) which generalizes  the Fluctuation
Dissipation Theorem (FDT) for a weakly out of equilibrium system
 (for a review see ref. \cite{Mezard,Cugliandolo,Peliti}). In order
to understand this generalization, we  recall the main
consequences of FDT in a system which is in thermodynamic
equilibrium. We consider an observable $V$ of such a system and
its conjugate  variable  $q$ . The response function
$\chi_{Vq}(\omega)$, at frequency $f=\omega / 2 \pi$, describes
the variation $\delta V (\omega)$ of $V$ induced by a perturbation
$\delta q (\omega)$ of $q$, that is $\chi_{Vq} (\omega)=\delta V
(\omega)/ \delta q(\omega)$. FDT relates the fluctuation spectral
density of $V$ to the response function $\chi_{Vq}$ and the
temperature T of the system:

\begin{equation}
S(\omega) = { 2 k_B \ T \over \pi \omega } {\it
Im}\left[\chi_{Vq}(\omega) \right] \label{FDR}
\end{equation}

where $S(\omega)=<|V(\omega)|^2>$ is the fluctuation spectral
density of $V$, $k_B$ is the Boltzmann constant, ${\it Im}\left[
\chi_{Vq}(\omega) \right]$ is the imaginary part of
$\chi_{Vq}(\omega)$. Textbook examples of  FDT  are Nyquist's
formula relating the voltage noise to the electrical resistance
and the Einstein's relation for Brownian motion relating the
particle diffusion coefficient  to the fluid viscosity
\cite{book}.

When the system is not in equilibrium FDT, that is eq.\ref{FDR},
may fail. Indeed theoretical works \cite{Kurchan} predict a
violation of eq.\ref{FDR} which has been observed  \cite{Mezard}
in  many numerical simulations
(\cite{Peliti},\cite{Parisi}-\cite{Berthier}) and in a few
experiments \cite{Grigera}-\cite{buisson}.

Because of the slow dependence on $t_w$ of the response functions,
it has been proposed to use  a  FDR which generalizes eq.\ref{FDR}
and which can be used to define
  an effective temperature $T_{eff}(\omega,t_{w})$ of the system \cite{Peliti}:

\begin{equation}
T_{eff} (\omega,t_w) = { S(\omega,t_w) \ \pi \omega \over  {\it
Im}\left[ \chi_{Vq}(\omega,t_w) \right] \ 2 k_B }
 \label{Teff}
\end{equation}

It is clear that if eq.\ref{FDR} is satisfied $T_{eff}=T$,
otherwise $T_{eff}$ turns out to be a decreasing function of $t_w$
and $\omega$. The physical meaning of eq.\ref{Teff} is that there
is a time scale (for example $t_w$) which allows to separate the
fast processes from the slow ones. In other words the low
frequency modes, such that $\omega t_w < 1$, relax towards the
equilibrium value much slower than the high frequency ones which
rapidly relax to the temperature of the thermal bath. Therefore it
is conceivable that the slow frequency modes keep memory of higher
temperatures for a long time and for this reason their temperature
should be higher than that of the high frequency ones. This
striking behavior has been observed in several numerical  models
of aging \cite{Peliti},\cite{Parisi}-\cite{Berthier}. Further
analytical and numerical studies of simple models show that
eq.\ref{Teff} is a good definition of temperature in the
thermodynamic sense \cite{Cugliandolo,Peliti}. In spite of the
large amount of theoretical studies there are only a few
experiments where FDR is studied in aging materials. The
experimental analysis of the dependence of $T_{eff}(\omega,t_w)$
on $\omega$ and $t_w$ is very useful to distinguish among
different models of aging because the FDT violations are model
dependent \cite{Peliti},\cite{Parisi}-\cite{Berthier}. Furthermore
the direct analysis of the noise signal allows one to understand
if the dynamics is either intermittent or continuous in time.

Recently, a few experiments have analyzed this problems  in real
materials \cite{Grigera}, \cite{Bellon,BellonD}, \cite{Herisson},
\cite{buisson}. The violation of FDT measured in an experiment on
a spin glass \cite{Herisson} seems to be in agreement with
theoretical predictions,  which were originally based on mean
field approach of spin glasses.  In contrast, experiments done on
dielectric measurements on glycerol \cite{Grigera}, colloidal
glasses\cite{Bellon,BellonD} and polymers \cite{buisson} present
only a qualitative agreement with theory.

The  previous  analysis of Laponite \cite{Bellon,BellonD} and of
polycarbonate\cite{buisson} was mainly based on the study of the
time evolution of the noise spectra which is surprisingly similar
in these two very different materials. In both cases the effective
temperature defined using eq.\ref{Teff} is huge and the persistent
time of the violation is extremely long. We have therefore
analyzed directly the time evolution of the noise signal in both
experiments and we find a strongly intermittent beahvior in both
materials.  In this paper we describe the results of this analysis
and we  want also  to point out the  common features observed in
the slow relaxation dynamics of these two materials.

The paper is organized as follows. In section 2 and 3 we recall
the main results of the  experiment on Laponite and polycarbonate
respectively. We also describe the  analysis performed on the time
evolution of the noise signal.  In section 4 we compare the
results of the two experiments and we discuss their relevance for
recent aging models. Conclusions are done at the end of the
section.

\section{ Laponite  electric properties}

\subsection{The experimental apparatus}

The Laponite \cite{Laponite}  solution is used as a conductive
liquid between the two golden coated electrodes of a cell (see
fig.1). It is prepared in a clean $\mathrm{N_2}$ atmosphere to
avoid $\mathrm{CO_2}$ and $\mathrm{O_2}$ contamination, which
perturbs the electrical measurements. Laponite particles are
dissolved at a concentration of $2.5 \%$ mass fraction in pure
water under vigorous stirring during $300s$. To avoid the
existence of any initial structure in the sol, we pass the
solution through a
 $1\mu m$ filter when filling our cell. This instant defines the origin
 of the aging time $t_w$ (the filling of the cell takes roughly two
minutes, which can be considered the maximum inaccuracy of $t_w$).
The sample is then sealed so that no pollution or evaporation of
the solvent can occur. At this concentration, the  light
scattering experiments show that Laponite \cite{Laponite}
structure functions are still evolving $500h$ after the
preparation \cite{Kroon}.  We only study the beginning of this
glass formation process.

\begin{figure}[!ht ]
 \centerline{\epsfxsize=0.8\linewidth \epsffile{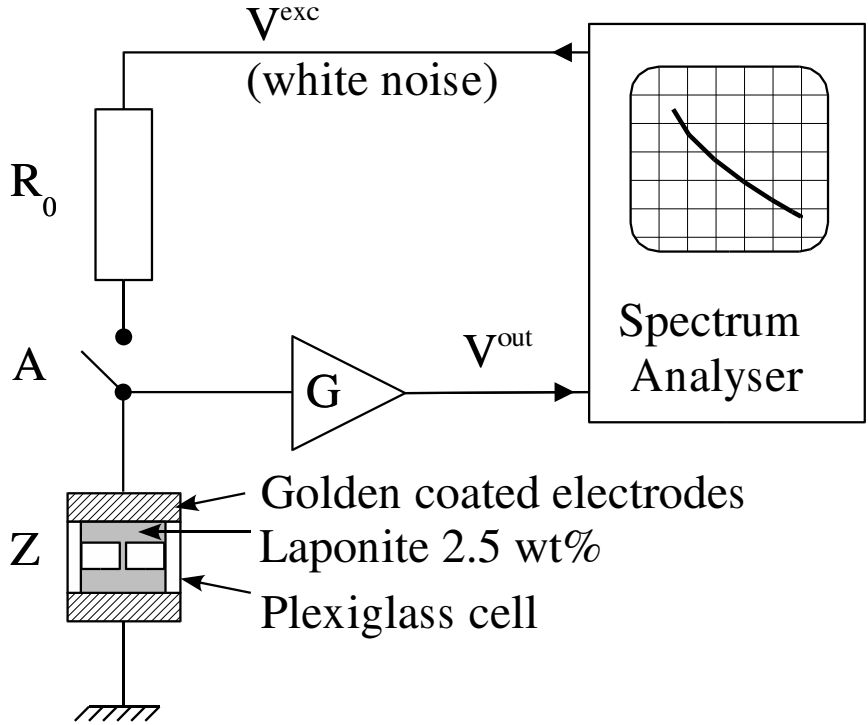}}
  \caption{{\bf  Laponite  experimental set-up} The impedance under test $Z$ is a
cell  filled with a $2.5 wt\%$ Laponite sol. The electrodes of the
cell are golden coated to avoid oxidation. One of the two
electrodes is grounded whereas the other is connected to the
entrance of a low noise  voltage amplifier characterized by a
voltage amplification $G$. With a spectrum analyzer, we
alternately record the frequency response
$FR(\omega)=<V^{out}/V^{exc}>$ (switch $A$ closed) and the
spectrum $S(\omega)=<|{V^{out}}|^2>$ (switch $A$ opened). The
input voltage $V^{exc}$ is a white noise excitation, thus from
$F\!R(\omega)$ we derive the impedance $Z(\omega)$ as a function
of $\omega$, that is $Z(\omega)=R_0 / (G / F\!R(\omega) - 1)$;
whereas from $S(\omega)$, we can estimate the voltage noise of
$Z$, specifically $S_Z(\omega) = [S(\omega)-S_a(\omega)]/G^2$
where $S_a(\omega)$ is the noise spectral density of the
amplifier} \label{measurement}
\end{figure}

The two electrodes of the cell are connected  to  our measurement
system, where we alternately record the cell electrical impedance
$Z(t_w,\omega)$ and the  voltage noise density $S_{Z}(t_w,\omega)$
(see fig.\ref{measurement}). Taking into account that  in this
configuration ${\it Im }\left[ \chi_{Vq}(t_w, \omega)
\right]=\omega {\it Re}\left[ Z(t_w, \omega) \right]$, one obtains
from eq.\ref{Teff} that the effective temperature of the Laponite
solution as a function of the aging time and frequency is:

\begin{equation}
T_{eff}(t_w,\omega)=\pi S_{Z}(t_w,\omega) / 2 k_B {\it Re}\left[
Z(t_w,\omega)\right]
 \label{TLap}
\end{equation}

which is an extension of the Nyquist formula.

The electrical impedance  of the sample is the sum of 2 effects:
the bulk is purely conductive, the ions of the solution follow the
forcing field, whereas the interfaces between the solution and the
electrodes give mainly a capacitive effect due to the presence of
the Debye layer\cite{hunter}. This beahvior has been validated
using a four-electrode potentiostatic technique \cite{electrochem}
to make sure that the capacitive effect is only due to the
surface. In order to test only bulk properties, the geometry of
the cell is tuned to push the surface contribution to low
frequencies. Specifically the cell is composed by two large
reservoirs in contact with the electrodes which have an area of
$25 cm^2$. The reservoirs are connected by a rigid tube (see
fig.\ref{measurement}) whose section and length give the main
contribution to the bulk electrical resistance. Thus by changing
the sizes of the tube the bulk resistance can be changed from $300
\Omega$ to $100K\Omega$. We checked that the dynamics of the
system does not depend on the value of the bulk resistance. For a
bulk resistance of about $10^5 \Omega$ the  cut-off frequency of
the equivalent R-C circuit (composed by the series of the Debye
layers plus the bulk resistance) is about $0.02Hz$. In other words
above this frequency the imaginary part of the cell impedance is
about zero.

\subsection{ FDR measurements }

\begin{figure}[!ht]
 % \centering
  {\bf (a)} \\
  \centerline{{\epsfysize=0.4\linewidth \epsffile{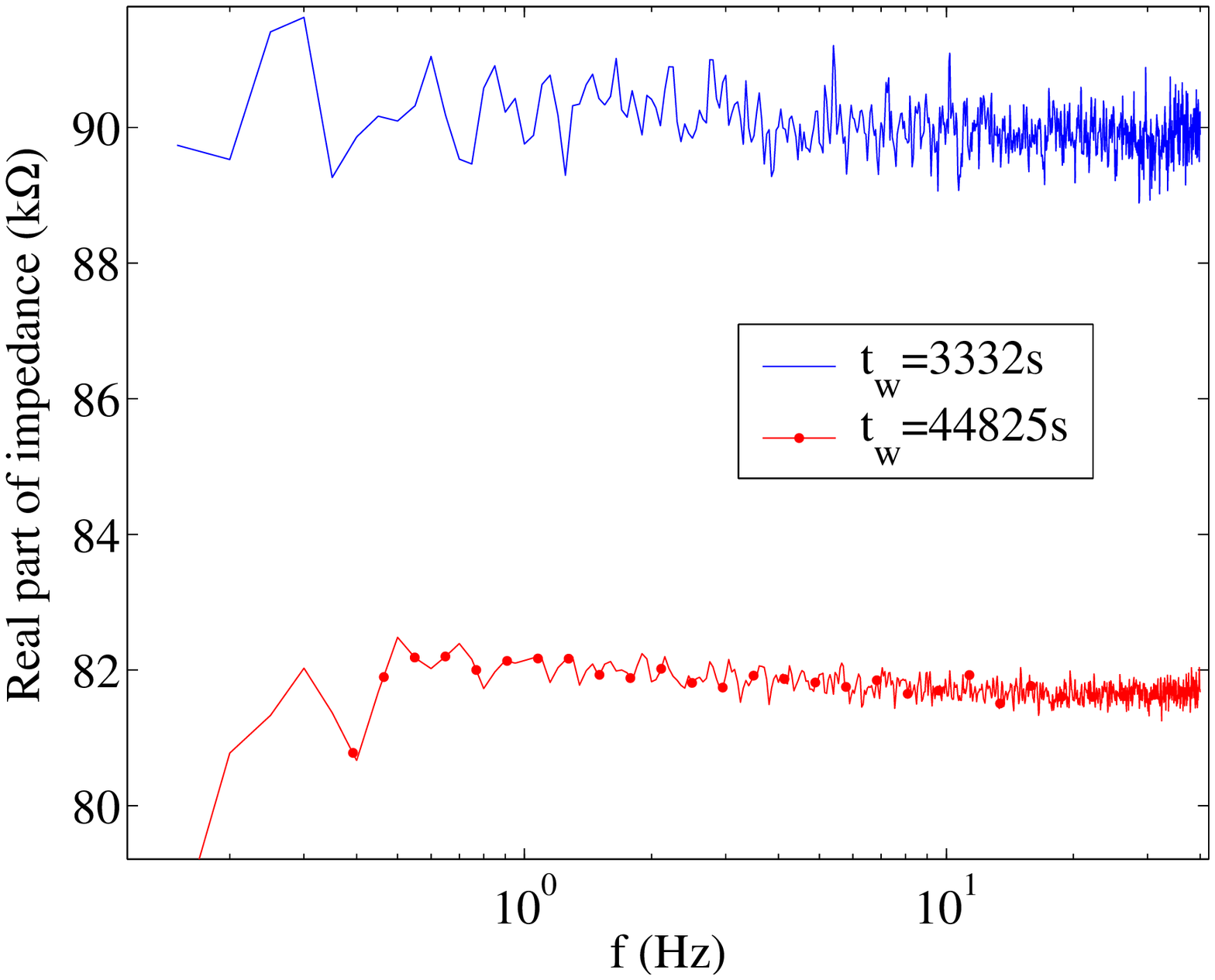}}}
  {\bf (b)} \\
  \centerline{\epsfysize=0.4\linewidth \epsffile{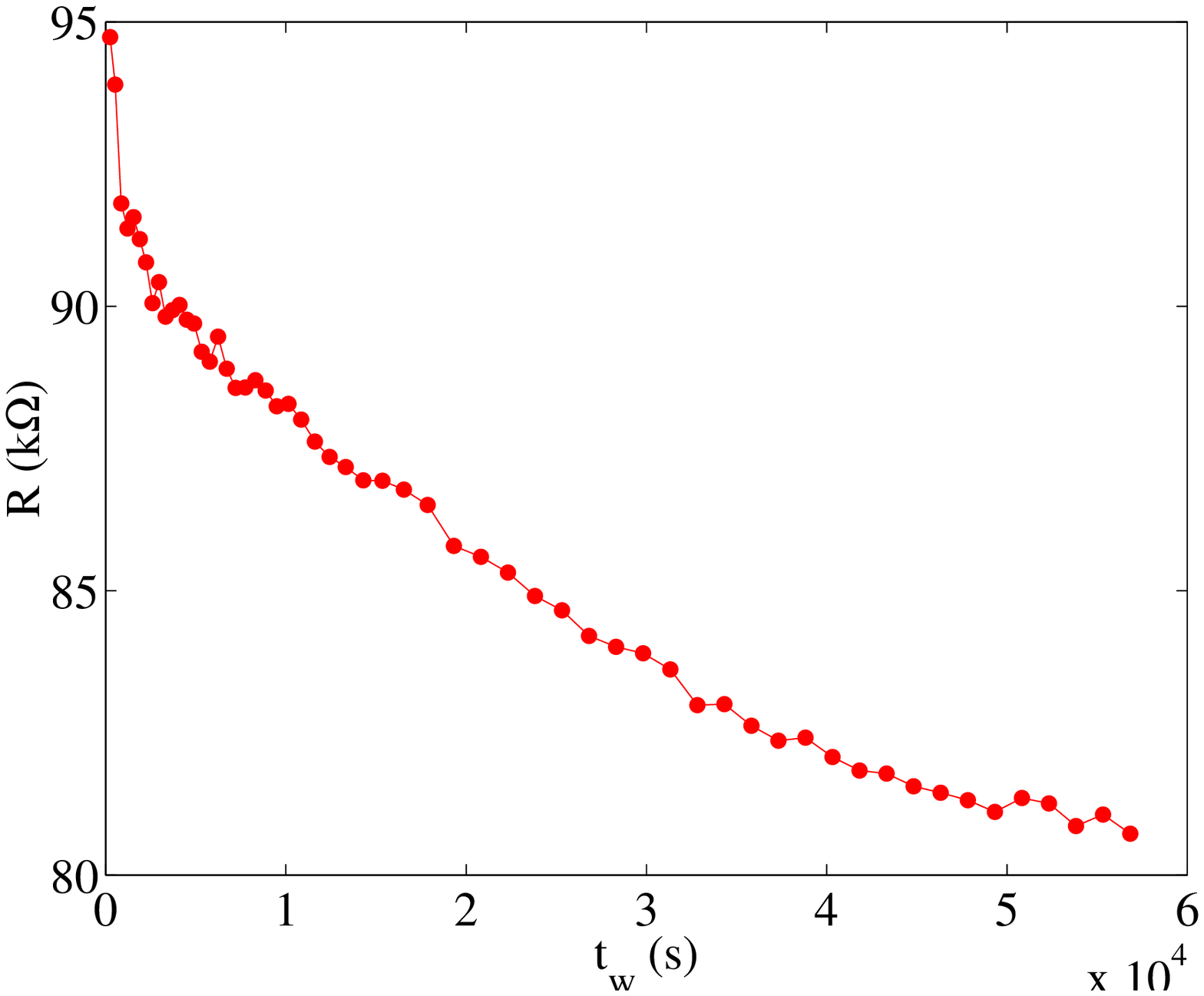}}
  \caption{{\bf  Laponite  response function} (a) Frequency dependence
  of a sample impedance for 2
different aging times: continuous line $t_w=3332s$ ; $ \ (\bullet)
t_w=44825s$.(b) Time evolution of the resistance. This long time
evolution is the signature of the aging of the sol. In spite of
the decreasing mobility of Laponite particles in solution during
the gelation, the electrical conductivity increases.}
\label{response}
\end{figure}

\begin{figure}[!h]
%\begin{center}
 \centerline{\epsfxsize=0.7\linewidth \epsffile{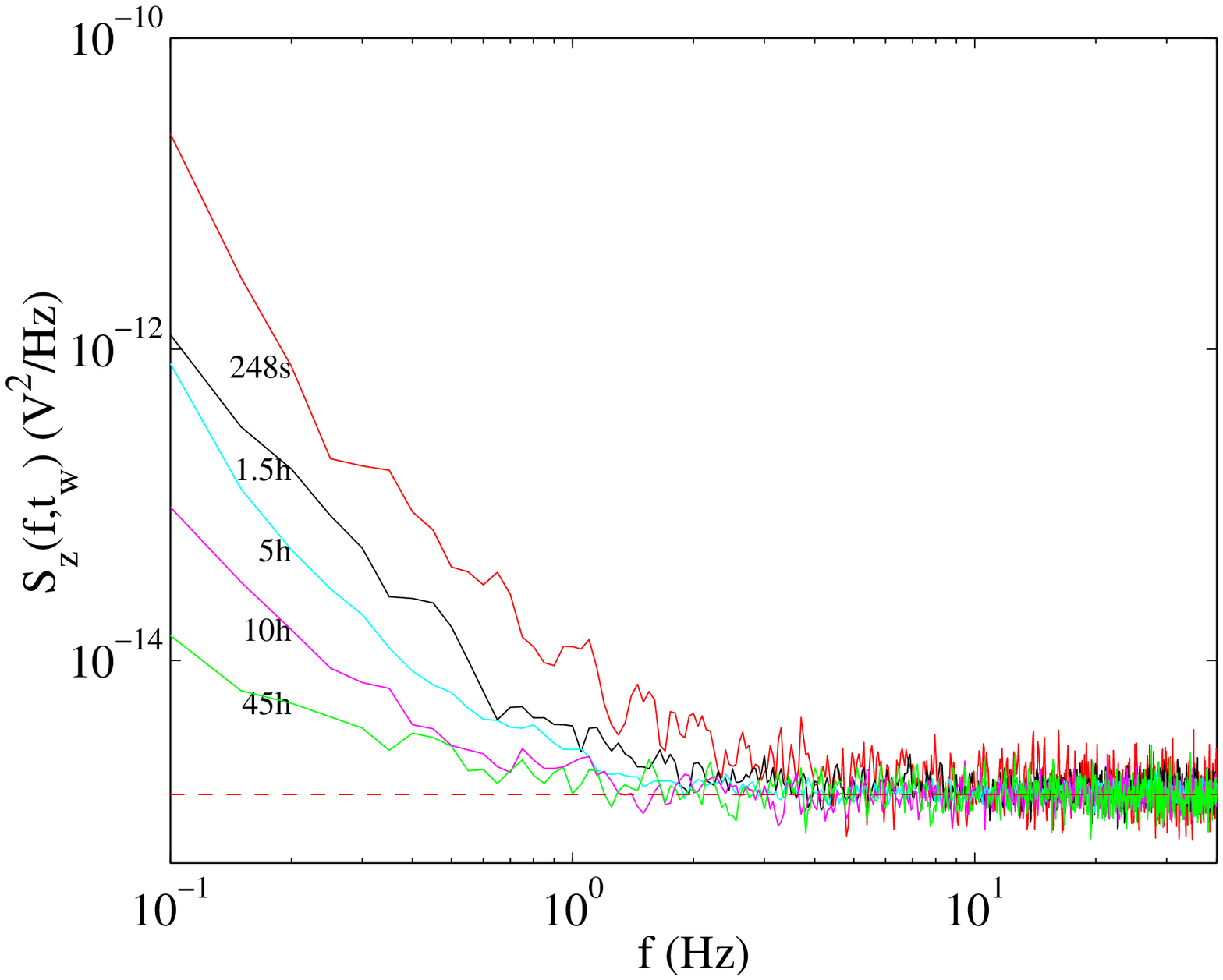}}
        %\psfrag{xl}[tc][tc]{\small Frequency $\nu$ ($Hz$)}
        %\psfrag{yl}[Bc][Bc]{\small $S_{Z}(t,\nu)$ ($V^2/Hz$)}
        %\psfrag{xle}[tc][tc]{\tiny $\omega (t/1h)^{0.5}$ ($rad.s^{-1}$)}
        %\psfrag{yle}[Bc][Bc]{\tiny $S_{Z}(t,f)$ ($V^2/Hz$)}
        %\psfrag{0.3h}[Br][Br]{\tiny $t=0.3\, h$}
        %\psfrag{1.5h}[Br][Br]{\tiny $1.5\, h$}
        %\psfrag{2.5h}[Br][Br]{\tiny $2.5\, h$}
        %\psfrag{5h}[Br][Br]{\tiny $5\, h$}
        %\psfrag{10h}[Br][Br]{\tiny $10\, h$}
        %\psfrag{50h}[Br][Br]{\tiny $50\, h$}
        %\psfrag{slope}[Bl][Bl]{\tiny slope $\omega^{-3.3}$}
       % \psfrag{3}[Br][Br]{ }
        %\includegraphics{Svzlionel.eps}
    %\end{center}
\caption{{\bf  Voltage fluctuations for Laponite } Voltage noise
density of one sample for different aging times. The horizontal
dashed line is the FDT prediction}\label{fluctuation}
\end{figure}

\begin{figure}[!ht]
 \centerline{\epsfxsize=0.7\linewidth \epsffile{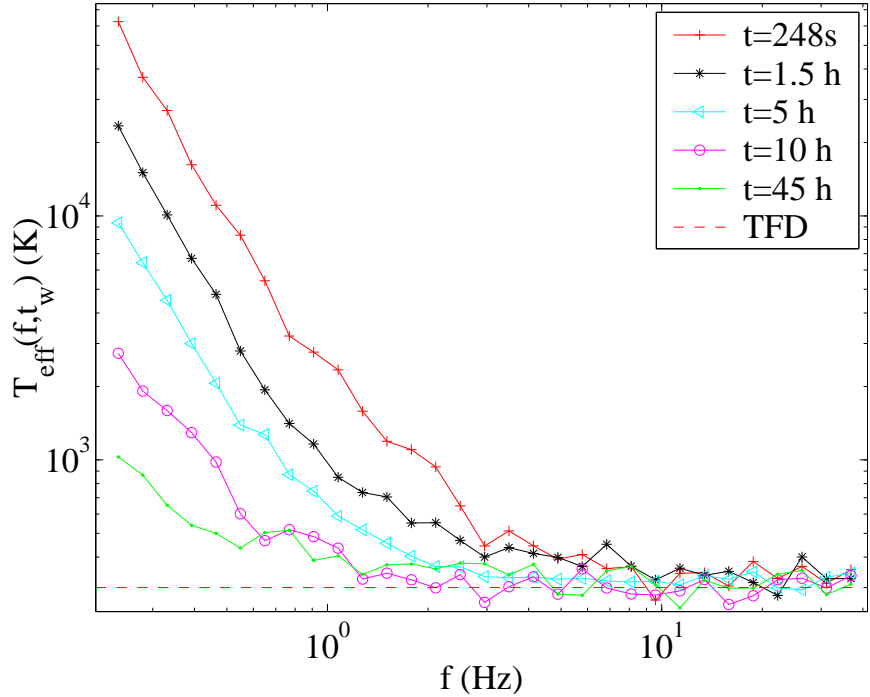}}
    %\begin{center}
     %   \psfrag{f}[cc][cc]{\small Frequency $\nu$ ($Hz$)}
      %  \psfrag{T}[cc][cc]{\small Effective temperature $T_{eff}$ ($K$)}
       % \psfrag{t=0.3 h}[Bl][Bl]{\tiny $t=0.3\, h$}
        %\psfrag{t=1.5 h}[Bl][Bl]{\tiny $t=1.5\, h$}
        %\psfrag{t=2.5 h}[Bl][Bl]{\tiny $t=2.5\, h$}
        %\psfrag{t=5 h}[Bl][Bl]{\tiny $t=5\, h$}
        %\psfrag{t=10 h}[Bl][Bl]{\tiny $t=10\, h$}
       % \psfrag{t=50 h}[Bl][Bl]{\tiny $t=50\, h$}
      %  \psfrag{TFD}[Bl][Bl]{\tiny TFD}
     %   \psfrag{7}[Br][Br]{ }
    %    \includegraphics{Tefflaponite.eps}
   % \end{center}
\caption{{\bf Effective temperature of Laponite}  Effective
temperature as a function of frequency for different aging times.
As $S_Z$ in fig.\ref{fluctuation}, $T_{ef\!f}$ strongly increases
and reaches huge values  for low frequencies and short aging
times.}\label{temperature}
\end{figure}

\begin{figure}[!ht]
 \centerline{\epsfxsize=0.5\linewidth \epsffile{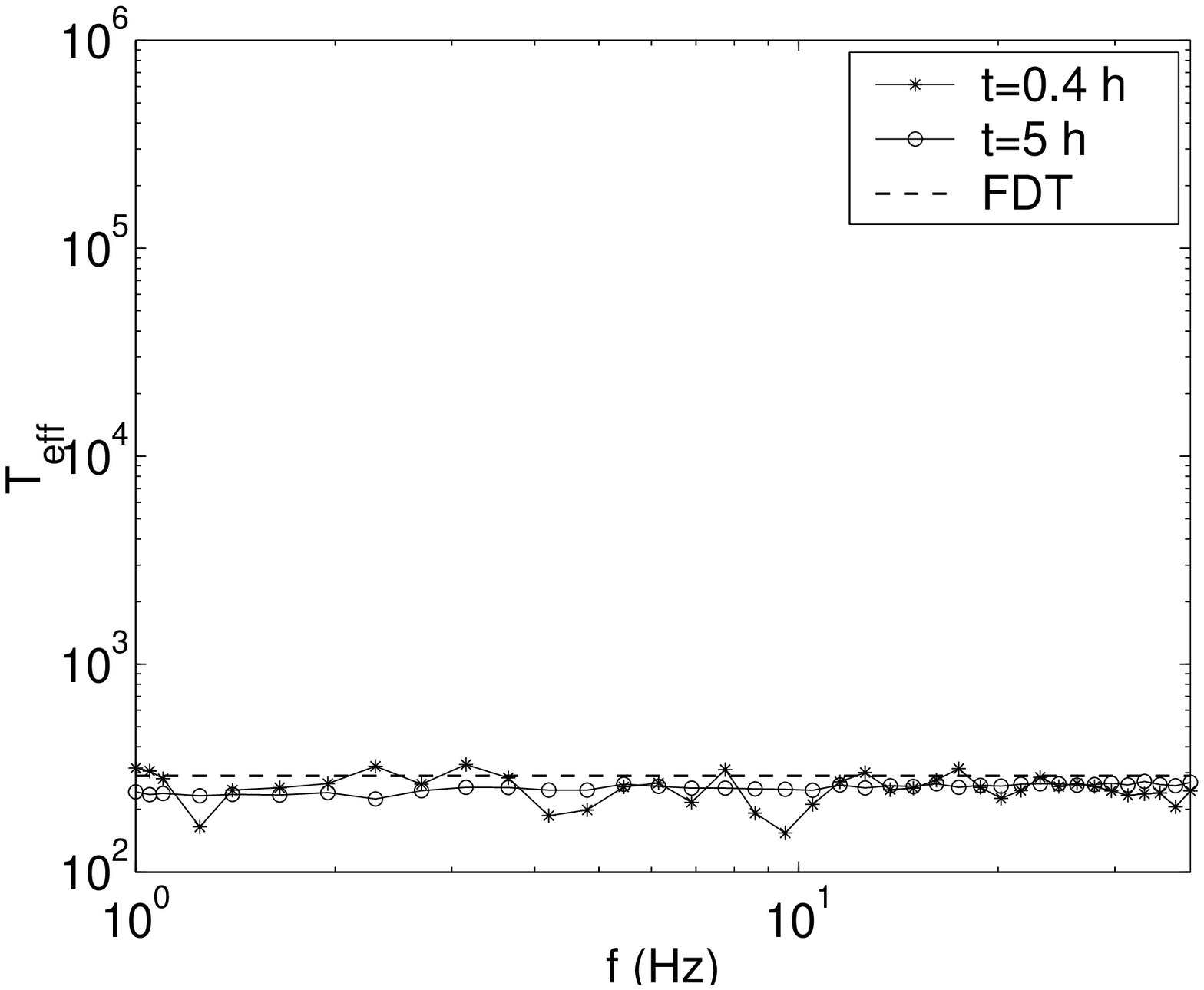}}
    %\begin{center}
     %   \psfrag{f}[cc][cc]{\small Frequency $\nu$ ($Hz$)}
      %  \psfrag{T}[cc][cc]{\small $T_{ef\!f}$ ($K$)}
       % \psfrag{t=0.4 h}[Bl][Bl]{\tiny $t=0.4\, h$}
        %\psfrag{t=5 h}[Bl][Bl]{\tiny $t=5\, h$}
       % \psfrag{TFD}[Bl][Bl]{\tiny TFD}
      %  \psfrag{7}[Br][Br]{ }
     %   \includegraphics{soude1.eps}
    %\end{center}
\caption{{\bf Effective temperature of an $NaOH$  solution in
water}. The effective temperature is plotted  as a function of
frequency for two different times after the preparation. This
solution has  a {\it p}H close to that of the Laponite,  and  no
violation is observed in this case  for any aging time.}
\label{fig:electrolyte}
\end{figure}

In fig.\ref{response}(a), we plot the real  part of the impedance
as a function of the frequency $f$, for a typical experiment and
two different times.  The time evolution of the resistance of one
of our sample is plotted in fig.\ref{response}(b): it is still
decaying in a non trivial way after $24 h$, showing that the
sample has not reached any equilibrium yet. This aging is
consistent with that observed in light scattering experiments
\cite{Kroon}.
As the dissipative part of the impedance $Re(Z)$ is weakly time
and frequency dependent, one would expect from the Nyquist formula
that so does the voltage noise density $S_{Z}$. But as shown in
fig.\ref{fluctuation}, FDR must be strongly violated for the
lowest frequencies and earliest times of our experiment: $S_{Z}$
changes by several orders of magnitude between highest values and
the high frequency tail \footnote{This low frequency noise cannot
be confused with the standard $1/f$ noise observed in many
electronic devices. We recall that the $1/f$ appears only when an
external current produced by an external potential  goes through
the device. In our cell no external potential is applied}. This
violation is clearly illustrated by the behavior of the effective
temperature in fig.\ref{temperature}\footnote{The usual
representation of the effective temperature in simulations is the
slope of the response versus correlation plot, but it is not
suited for our experimental data: the system being almost only
dissipative, the response function is close to a delta
distribution, thus FDR is only one point in this representation.}.
For long times and high frequencies, the FDR holds and the
measured temperature is the room one ($300K$); whereas for early
times $T_{eff}$ climbs up to $3.10^3K$ at $1Hz$.  Moreover,
$T_{eff}$ could  be even larger for lower frequencies and lower
aging  times:  indeed, we found in all the tested samples no
evidence of a saturation of this effective temperature in our
measurement range. In order to be sure that the observed violation
is not due to an artifact of the experimental procedure, we filled
the cell with an electrolyte solution with {\it p}H close  to that
of the Laponite sol such that the electrical impedance of the cell
was the same. Specifically we filled the cell with $NaOH$ solution
in water at a concentration of $10^{-3} \ mol.l^{-1}$. The results
of the measurements of $T_{eff}$ are shown  in
fig.\ref{fig:electrolyte} at two different times after the sample
preparation.   In this case we did not observe any violation of
FDR at any time.

%\newpage

\newpage

\subsection{ Statistical analysis of the noise}

\begin{figure}[!ht]
\centerline{\bf \large (a) \hspace{9cm} (b)}
    \begin{center}
        \includegraphics[width=8cm]{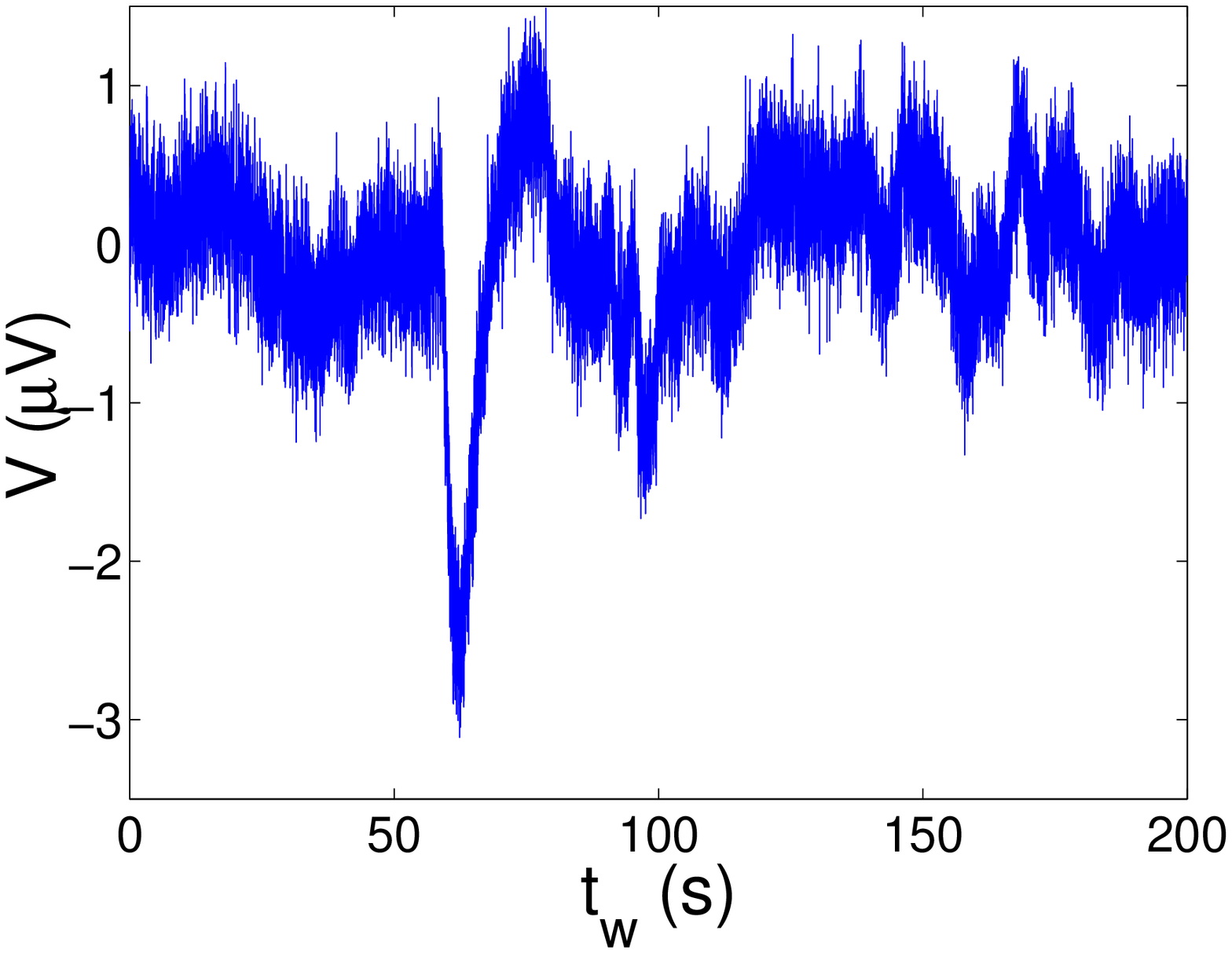}
        \hspace{1mm}
        \includegraphics[width=8cm]{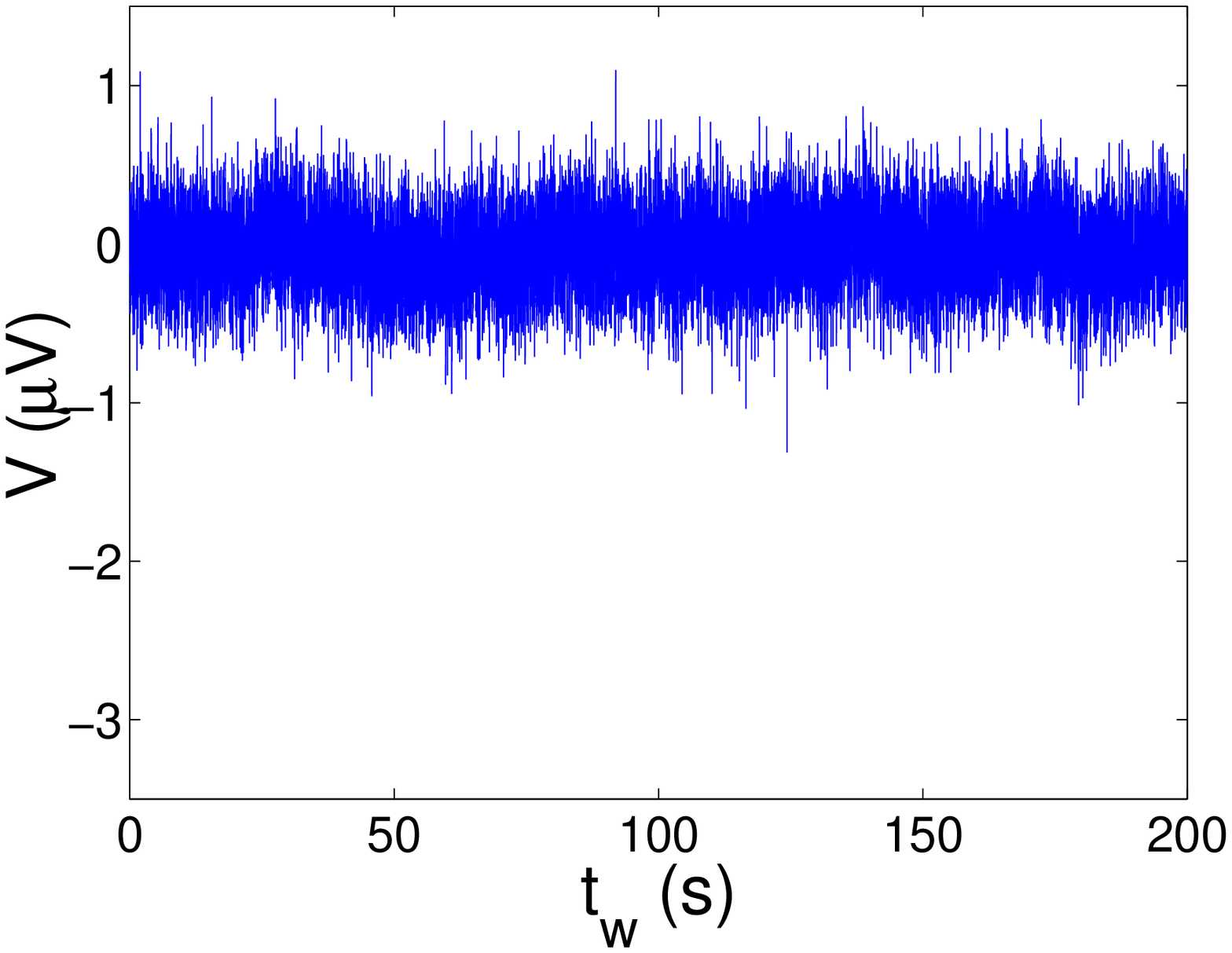}
    \end{center}
\caption{{\bf  Voltage noise signal in Laponite }. (a) Noise
signal, 2 hours after the Laponite preparation, when FDT is
violated.  (b) Typical noise signal when FDT is not violated. }
\label{fig:signal}
\end{figure}

\begin{figure}[!ht]
    \begin{center}
        \includegraphics[width=10cm]{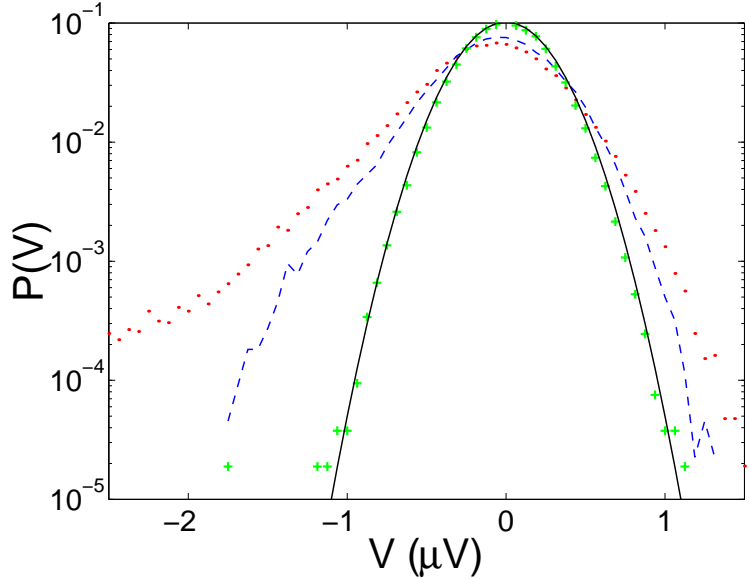}
        \end{center}
\caption{{\bf PDF of the  voltage noise in Laponite }. Typical PDF
of the noise signal at different times after preparation, with
from top to bottom: $
 \ (...) t_w=1h, \ (- -) t_w=2h, \ (+)
t_w=50h$. The continuous line is obtained from the FDT prediction.
} \label{fig:PDFlaponite}
\end{figure}

In order to understand such a beahvior  we have directly analyzed
the noise  voltage  across the Laponite cell. This test can be
safely done in our experimental apparatus because the amplifier
noise is negligible with respect to the thermal noise of the
Laponite cell even when FDT is satisfied. In
fig.\ref{fig:signal}(a) we plot a typical signal measured $2h$
after the gel preparation when the FDT is strongly violated. The
signal plotted in fig.\ref{fig:signal}(b) has been measured when
the system is relaxed and FDT is satisfied in all the frequency
range. By comparing the two signals we immediately realize that
there are very important differences. The signal in
fig.\ref{fig:signal}(a) is interrupted by bursts of very large
amplitude  which are responsible for the increasing of the noise
in the low frequency spectra (see fig.\ref{fluctuation}). The
relaxation time of the bursts has no particular meaning, because
it corresponds just to the characteristic time of the filter used
to eliminate the very low frequency trends.   As time goes on, the
amplitude of the bursts reduces and the time between two
consecutive bursts becomes longer and longer. Finally they
disappear as can be seen  in the signal  of
fig.\ref{fig:signal}(b) recorded after $50h$ when the system
satisfies FDT.
The evolution of the intermittent properties of the noise  can be
characterized by studying the probability density function(PDF) of
the signal as a function of time. To compute the PDF,  the time
series are  divided in several time windows and the PDF are
computed in each of these window. Afterwards the result of several
experiments are averaged. The PDF computed at different times are
plotted in fig.\ref{fig:PDFlaponite}. We see that at short $t_w$
the PDF presents very high tails which slowly disappear at longer
$t_w$. Finally a Gaussian shape is recovered at $t_w=16h $. This
kind of  evolution of the PDF clearly indicate that the signal is
very intermittent at the very beginning and it relaxes to the
Gaussian noise at very long times.

The comparison of these results with aging models will be done in
the conclusions. We prefer to describe  now  another experiment in
a completely different material.

\section{ Polycarbonate dielectric properties}

In order to give more insight into the problem of the violation of
FDT and of the intermittent beahvior discussed in the previous
section  we have done wide band ($20mHz-100Hz$) measurements of
the dielectric susceptibility and of  the polarization noise in a
polymer glass: polycarbonate. We present in this article several
results which show a strong violation of the FDT  when this
material is  quenched from the molten state to below its
glass-transition temperature. The effective temperature defined by
eq.\ref{Teff} slowly relaxes towards the bath temperature. The
violation is observed even at  $\omega t_w \gg 1$  and it may last
for more than $3h$ for $f>1Hz$.

\subsection{The experimental apparatus}

The polymer used in this investigation is Makrofol DE 1-1 C, a
bisphenol A polycarbonate, with $T_g \simeq 419K$, produced by
Bayer in form of foils. We have chosen this material because it
has a wide temperature range of strong aging \cite{Struick}. This
polymer  is totally amorphous: there is no evidence of
crystallinity \cite{Wilkes1}. Nevertheless, the internal structure
of polycarbonate changes and relaxes as a result of a change in
the chain conformation by molecular
motions\cite{Struick},\cite{Duval},\cite{Quinson}. Many studies of
the dielectric susceptibility of this material exist, but no one
had an interest on the problem of noise measurements.

\begin{figure}[!ht]
\begin{center}
\includegraphics[width=9.5cm]{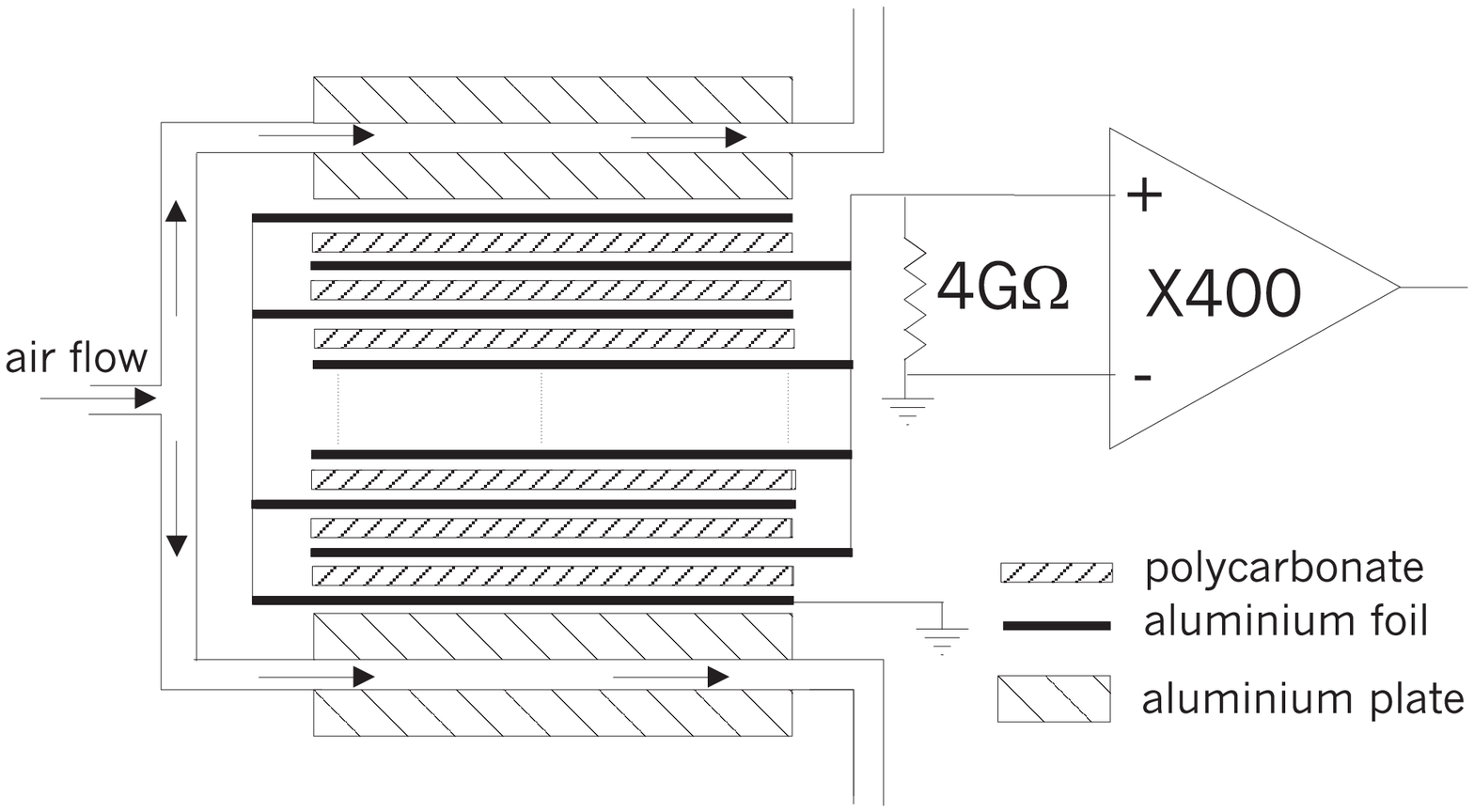}
        \hspace{1mm}
        \includegraphics[width=6cm]{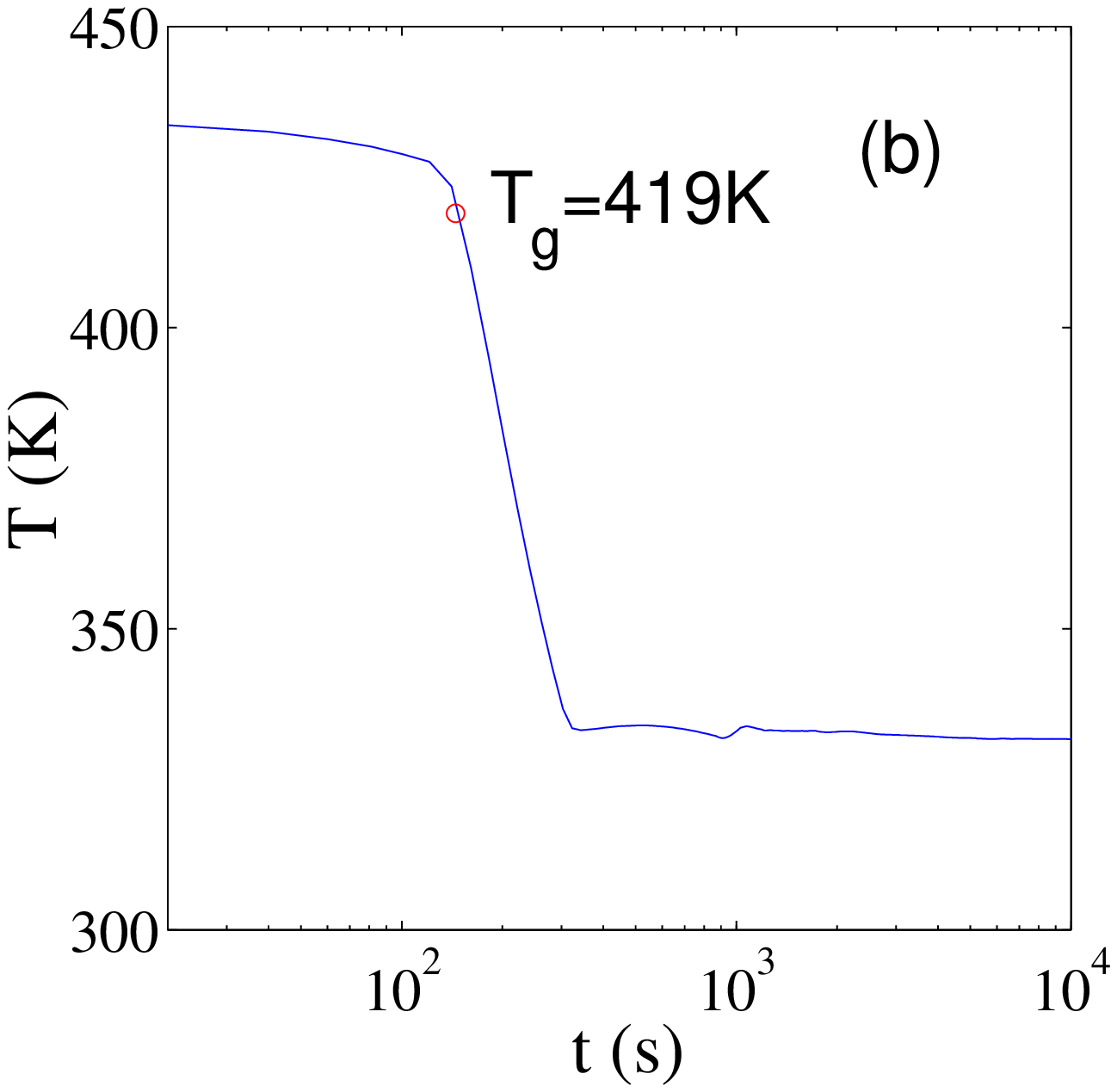}
\end{center}
\caption{(a) {\bf  Polycarbonate experimental  set-up} (b)  {\bf
Typical temperature quench:} from $T_i=433K$ to $T_f=333K$, the
origin of $t_w$  is at $T=T_g$. }
 \label{Experimental set-up}
\end{figure}

In our experiment polycarbonate is  used as the dielectric of a
capacitor. The  capacitor is composed by $14$ cylindrical
capacitors in parallel in order to reduce the resistance of the
sample and to increase its capacity. Each capacitor  is made of
two aluminum electrodes, $12\mu m$ thick, and  by  a disk of
polycarbonate of diameter $12cm$ and thickness $125\mu m$. The
experimental set-up is shown in fig.\ref{Experimental set-up}a).
The $14$ capacitors are sandwiched together and put inside two
thick aluminum plates which contain an air circulation used  to
regulate the sample temperature.  This mechanical design of the
capacitor is very stable and gives very reproducible results even
after many temperature quenches. The capacitor is inside two
Faraday screens to insulate it from external noise. The
temperature of the sample is controlled within a few percent. Fast
quench of about $50K/min$ are obtained by injecting Nitrogen vapor
in the air circulation of the aluminum plates. The electrical
impedance of the capacitor is $Z(\omega,t_w) = R / (1+i \omega \ R
\ C)$, where $C$ is the capacitance and $R$ is a parallel
resistance which accounts for the complex dielectric
susceptibility. It is measured using a Novocontrol Dielectric
Analyzer. The noise spectrum of this impedance $S_Z(\omega,t_w)$
is:
\begin{equation}
S_Z(f,t_w)= 4 \ k_B \ T_{eff}(f, t_w) \  Re  [Z(\omega,t_w)]= {4 \
k_B \ T_{eff}(f, t_w) \ R \over 1+ (\omega \ R \ C)^2 } \label{SZ}
\end{equation}
where $T_{eff}$ is the effective temperature of the sample.
 In order to measure $S_Z(f,t_w)$,
  we have made a differential amplifier based on selected low
noise JFET(2N6453 InterFET Corporation), whose input has been
polarized by a resistance $R_i= 4G\Omega$. Above $2Hz$, the input
voltage noise of this amplifier is $5nV/\sqrt{Hz}$ and the input
current noise is about $1fA/\sqrt{Hz}$. The output signal of the
amplifier is analyzed either by an HP3562A dynamic signal analyzer
or directly acquired by a NI4462 card. It is easy to show that the
measured spectrum at the amplifier input is:
\begin{eqnarray}
S_V(f,t_w)& = &{4 \  k_B \ R \ R_i \ \ (\ T_{eff}(f, t_w) \ R_i +
\ T_R \ R +  S_\xi(f) \ R \ R_i ) \over (R+R_i)^2+(\omega \ R \
R_i \ C)^2} + S_{\eta}(f)
%= \\ \notag
%\\
%& = &{ S_Z(f,t_w) \ R_i^2 \ [ 1+ (\omega \ R \ C)^2] + \ 4 \  k_B
%\ T \ R^2 \ R_i  + S_\xi(f) \ (R \ R_i)^2 \over (R+R_i)^2+(\omega
%\ R \ R_i \ C)^2} + S_{\eta}(f)
 \label{Vnoise}
\end{eqnarray}
where $T_R$ is the temperature of $R_i$ and $S_\eta$ and $S_\xi$
are respectively the voltage and the current noise spectrum of the
amplifier. In order to reach the desired statistical accuracy  of
 $S_V(f,t_w)$, we averaged the results of many
 experiments. In  each of these experiments
 the sample is first heated  to $T_i=433K$. It is   maintained  at this temperature
   for 4 hours in order to
reinitialize its thermal history. Then it is quenched from $T_i$
to $T_f=333K$ in about 2 minutes. A typical thermal history of the
quench is shown in fig.\ref{Experimental set-up}(b). The
reproducibility of the capacitor impedance, during this thermal
cycle   is always better than $1\%$. The origin of aging time
$t_w$ is the instant when the capacitor temperature is at $T_g
\simeq 419 K$, which of course may depend on the cooling rate.
However adjustment of $T_g$ of a few degrees will shift the time
axis by at most $30s$, without affecting our results.

\subsection{FDR measurements}

\begin{figure}[!ht]
 % \centering
  %{\bf (a)} \\
  \centerline{\epsfysize=0.45\linewidth \epsffile{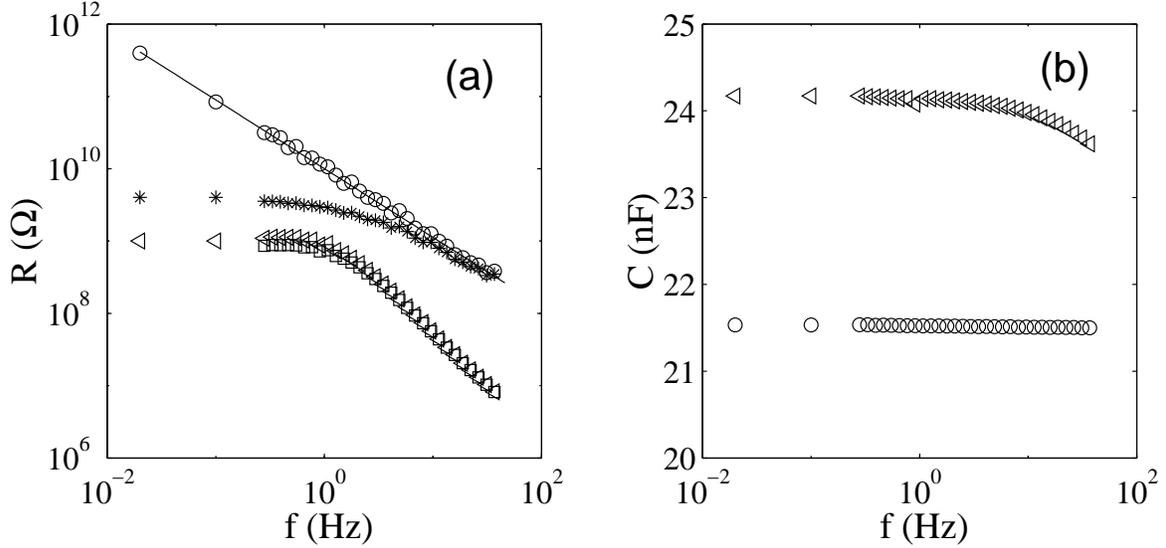}}
  %{\bf (b)} \\
  %\centerline{\epsfysize=0.5\linewidth \epsffile{imagimpedance.eps}}
  \caption{ {\bf Polycarbonate response function} (a) Polycarbonate resistance $R$
  as  a function of frequency measured at $T_i=433K$ ($\vartriangleleft$) and
  at  $T_f=333k$ ($\circ$). The effect of the $4G\Omega$ input resistance is also shown at $T=433K$ ($\square$) and
   at $T=333K$ ($\ast$).
   (b)  Polycarbonate capacitance versus frequency measured at $T_i
=433K$ ($\vartriangleleft$) and
  at  $T_f=333k$ ($\circ$).}
%c) Typical aging of $R$ measured at $1Hz$ as a function of $t_w$
%d) Typical temperature quench from $T_i=433K$ to $T_f=333K$, the
%origin of $t_w$  is at $T=T_g$.}
\label{reponse}
  \end{figure}

In fig.\ref{reponse}(a) and (b), we plot the measured values of
$R$ and $C$
 as a function of $f$ at $T_i$ and at $T_f$ for $t_w \geqslant 200s$.
  We see that lowering temperature $R$ increases and $C$
decreases. At $T_f$ aging is small and  extremely slow. Thus for
$t_w>200s$ the impedance can be considered constant without
affecting our results.  From the data plotted in fig.\ref{reponse}
(a) and (b) one finds that $R=10^{10}(1 \pm 0.05) \ f^{-1.05\pm
0.01} \ \Omega$ and $C=(21.5 \pm 0.05) nF$. In
fig.\ref{reponse}(a) we also plot the total resistance at the
amplifier input which is the parallel of the capacitor impedance
with $R_i$. We see that at $T_f$ the input impedance of the
amplifier is negligible  for $f>10Hz$, whereas it has to be taken
into account at slower frequencies.

%%%%%%%%%%%% FIGURE REPONSE %%%%%%%%%%%%%%%%%%%%%%%%%%%%%%%%%%%
\begin{figure}[!ht]
{\bf \hspace{20mm}(a)}

\centerline{\epsfxsize=0.6\linewidth\epsffile{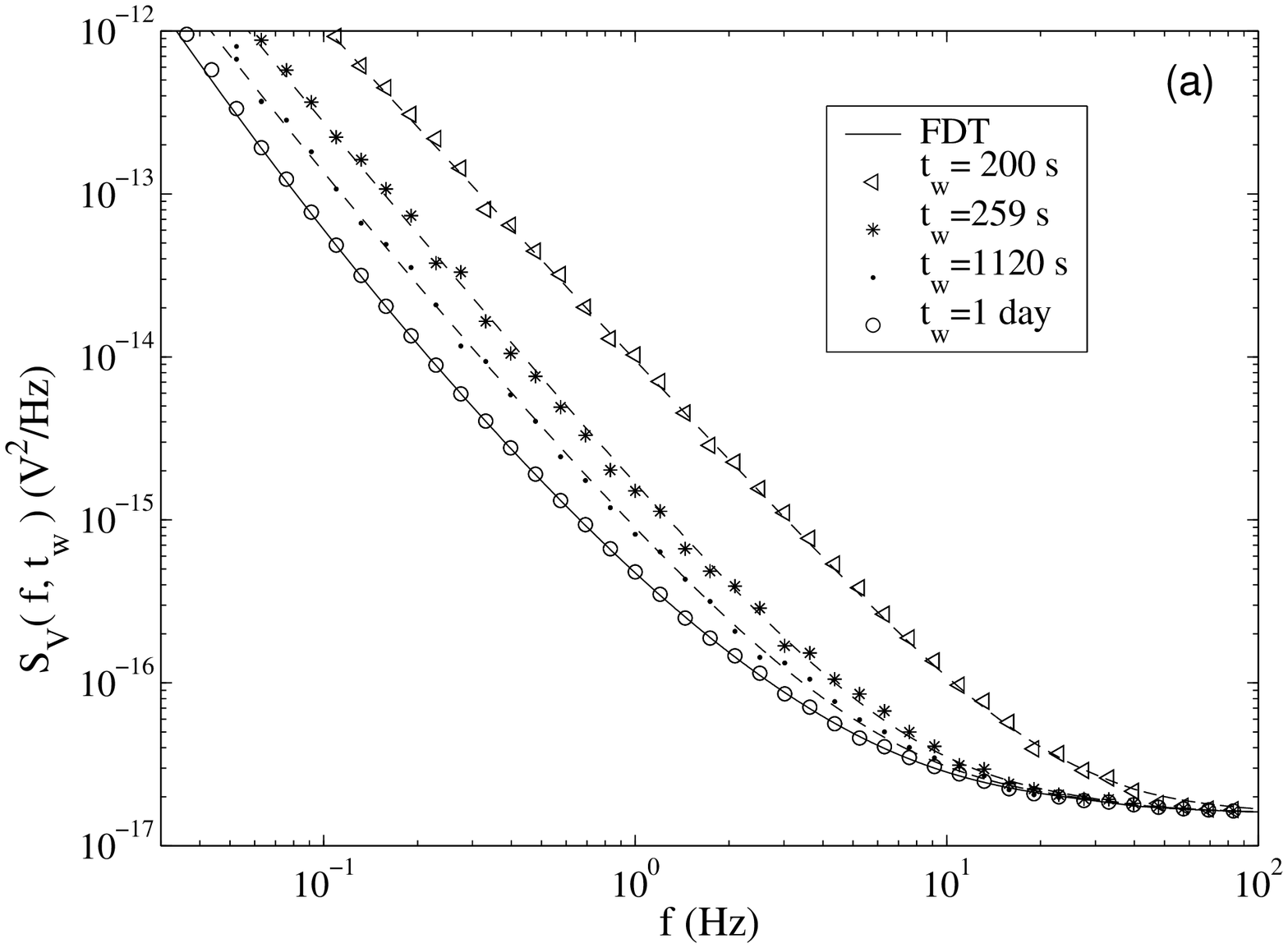} }

{\bf \hspace{20mm}(b)}

\centerline{\epsfxsize=0.6\linewidth\epsffile{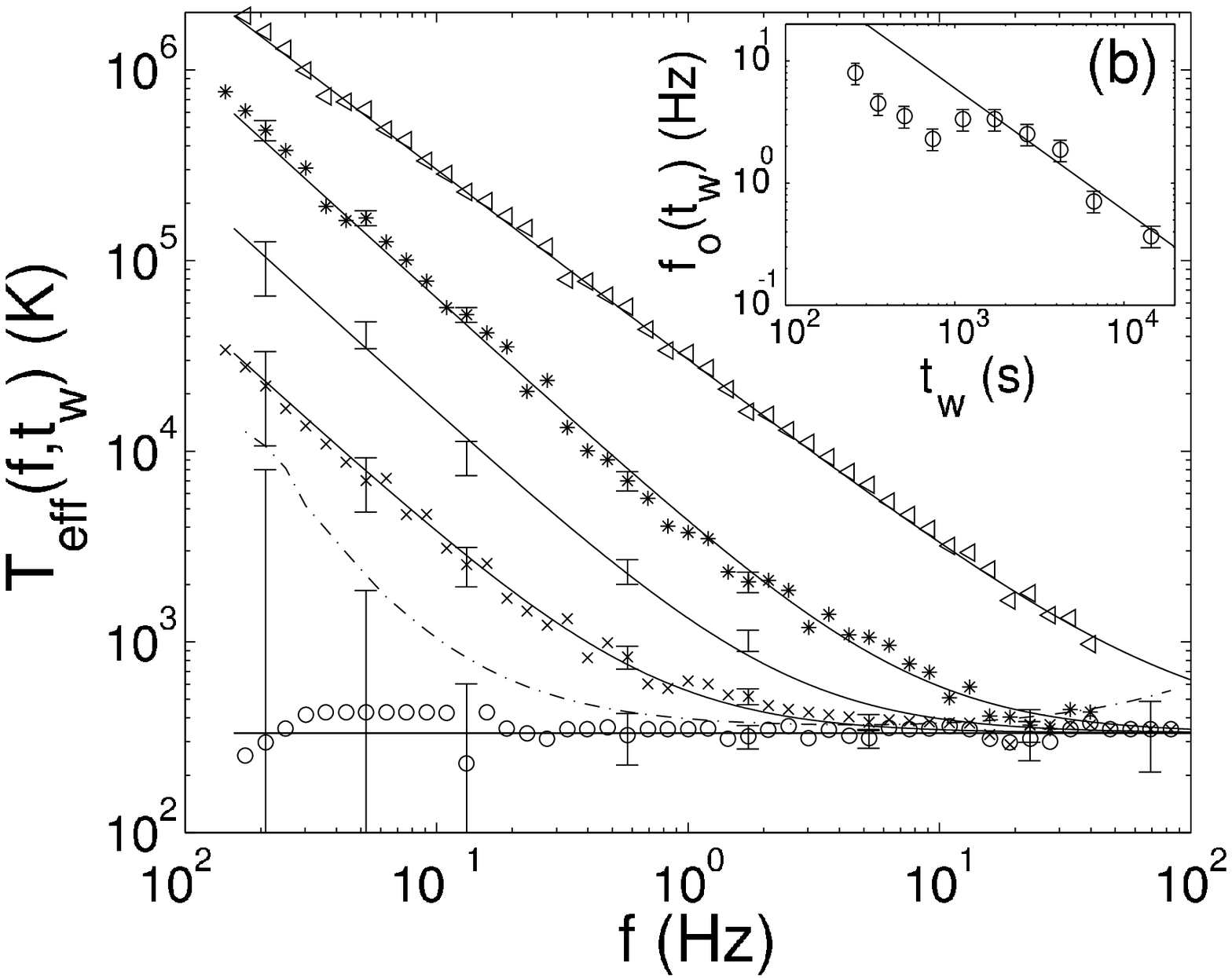}}

  \caption{ {\bf Voltage noise and effective temperature in polycarbonate}
   (a)  Noise power spectral density $S_V(f,t_w)$ measured at
$T_f= 333K$ and different
  $t_w$. The spectra are the average over seven quenches.
  The continuous line is the FDT prediction. Dashed lines
   are the fit obtained using eq.\ref{Vnoise} and eq.\ref{fitTeff} (see
   text for details). (b)  Effective temperature vs frequency at
$T_f=333K$ for different aging times: $ (\vartriangleleft)\ tw=
200 \ s$, $ (\ast)\ tw= 260 s$,  $ \bullet \ tw= 2580 s$, $
(\times) t_w=6542s$, $ (\circ) t_w= 1\ day $. The continuous lines
are the fits obtained using eq.\ref{fitTeff}. The horizontal
straight line is the FDT prediction.  The dot dashed line
corresponds to the limit where the FDT violation can be detected.
In the inset the frequency  $f_o(t_w)$, defined in
eq.\ref{fitTeff},is plotted as a function of $t_w$. The continuous
line is not a fit, but it corresponds to $f_o(t_w) \propto 1/t_w
$. } \label{noise}
\end{figure}

%%%%%%%%%%%%%%%%%%%%%%%%%%%%%%%%%%%%%%%%%%%%%%%%%%%%%%%%%%%%%%%%%%%%%%%%%%%
 fig.\ref{noise}(a) represents the evolution of
$S_V(f,t_w)$ after a quench. Each spectrum is obtained as an
average in a time window starting at $t_w$. The time window
increases with $t_w$ so to reduce error for large $t_w$. Then the
results of 7 quenches have been averaged. The longest time ($t_w=1
\  day$) the equilibrium FDT prediction (continuous line) is quite
well satisfied. We clearly see that FDT is strongly violated for
all frequencies at short times. Then high frequencies relax on the
FDT, but there is a persistence of the violation for lower
frequencies. The amount of the violation can be estimated by the
best fit  of $T_{eff}(f,t_w)$ in eq.\ref{Vnoise} where all other
parameters are known. We started at very large $t_w$ when the
system is relaxed and $T_{eff}=T$ for all frequencies. Inserting
the values in eq.\ref{Vnoise} and using the $S_V$ measured at
$t_w=1 day s$ we find $T_{eff}\simeq 333K$, within error bars for
all frequencies (see fig.\ref{noise}b). At short $t_w$  data show
that $T_{eff}(f,t_w)\simeq T_f$ for $f$ larger than a cutoff
frequency $f_o(t_w)$ which is a function of $t_w$. In contrast,
for $f<f_o(t_w)\ \ $ we find that $T_{eff}$ is:
$T_{eff}(f,t_w)\propto f^{-A(t_w)}$, with $A(t_w)\simeq 1$. This
frequency dependence of $T_{eff}(f,t_w)$ is quite well
approximated by

\begin{equation} T_{eff}(f,t_w)= T_f \ [ \ 1 \ + \ ( {f \over
f_o(t_w)})^{A(t_w)} \ ]
 \label{fitTeff}
\end{equation}

where $A(t_w)$ and  $f_o(t_w)$  are the  fitting parameters. We
find that $1<A(t_w)<1.2$ for all the data set. Furthermore  for
$t_w \geq 250$, it is enough to keep $A(t_w)=1.2$ to fit the data
within error bars. For $t_w <250s$ we fixed $A(t)=1$. Thus the
only free parameter in eq.\ref{fitTeff} is  $f_o(t_w)$.  The
continuous lines in fig.\ref{noise}(a) are the best fits of $S_V$
found inserting eq.\ref{fitTeff} in eq.\ref{Vnoise}.

 In fig.\ref{noise}(b) we plot the estimated $T_{eff}(f,t_w)$  as a
function of frequency at different $t_w$. We see that just after
the quench  $T_{eff}(f,t_w)$ is much larger than $T_f$  in all the
frequency interval. High frequencies rapidly decay towards the FDT
prediction whereas  at the smallest frequencies $T_{eff}\simeq
10^5K$. Moreover we notice that low frequencies decay more slowly
than high frequencies and that the evolution of $T_{eff}(f,t_w)$
towards the equilibrium value is very slow.
 From the data of fig.\ref{noise}(b) and eq.\ref{fitTeff}, it is
easy  to see that $T_{eff}(f,t_w)$ can be superposed onto a master
curve by plotting them as a function of $f/f_o(t_w)$. The function
$f_o(t_w)$ is a decreasing function of $t_w$, but the dependence
is not a simple one, as it can be seen in the inset of
fig.\ref{noise}(b). The continuous straight line is not fit,  it
represents $f_o(t_w)\propto 1/t_w$ which seems a reasonable
approximation  for these data. For $t_w > 10^4 s$ we find the
$f_o<1Hz$. Thus we cannot follow the evolution of $T_{eff}$
anymore  because the contribution of the experimental noise on
$S_V$ is too big, as it is shown in fig.\ref{noise}(b) by the
increasing of the error bars for  $t_w=1 \ day$  and $f<0.1 Hz$.

\begin{figure}[!ht]
{\bf \hspace{20mm} (a)}

  \centerline{\epsfxsize=0.6\linewidth \epsffile{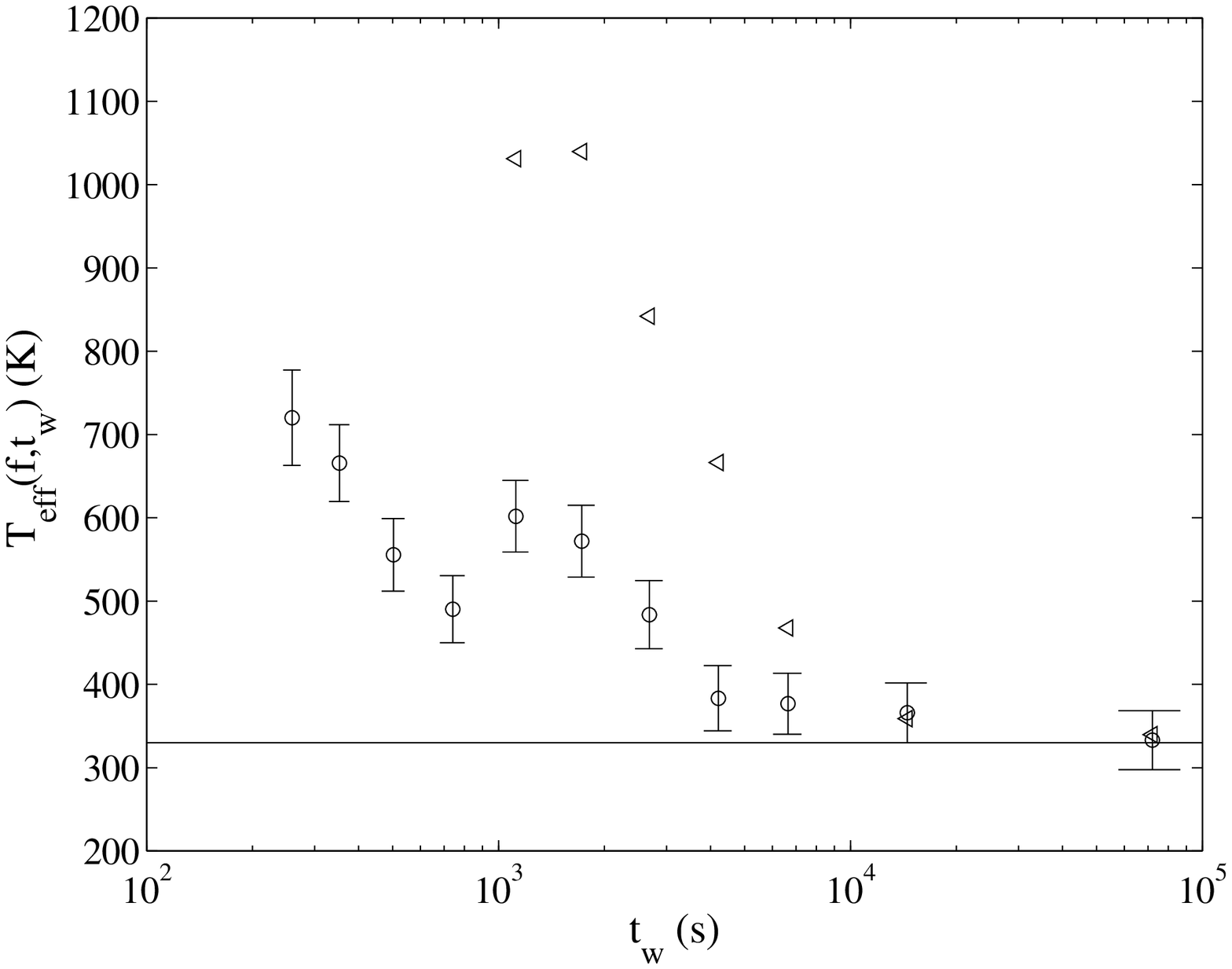}}

{\bf  \hspace{20mm} (b)}

 \centerline{\epsfxsize=0.6\linewidth \epsffile{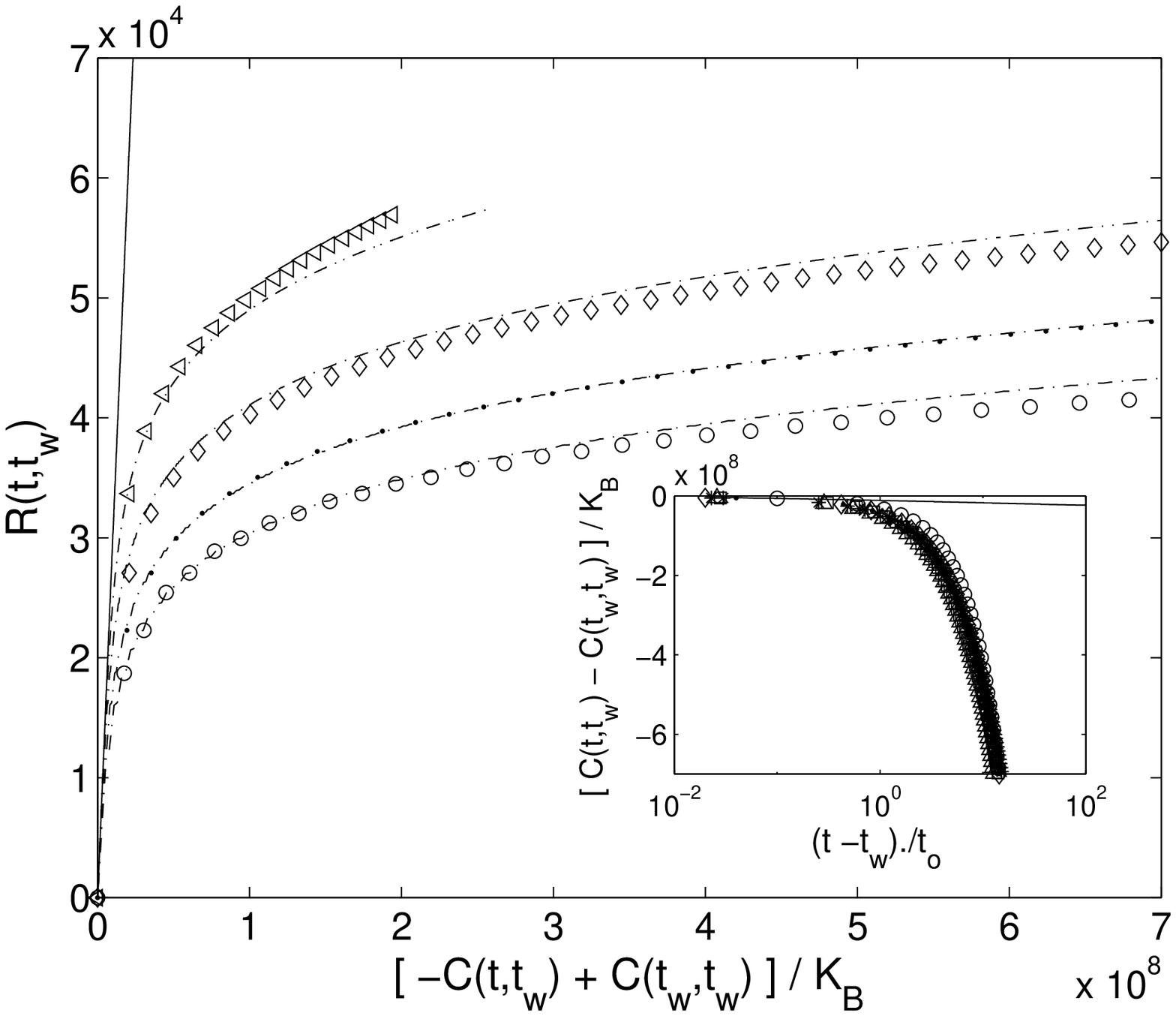}}
  \caption{  {\bf Violation of FDT in polycarbonate}  (a)  Effective
   temperature at $7Hz \ (\circ)$ and $2 Hz \ (\vartriangleleft)$
   measured as a function of $t_w$ at $T_f= 333K$. b) Plot of the
    integrated response $R(t,t_w)$ as a function of
$-C(t,t_w)+C(t_w,t_w)$ at different
  $t_w$. Symbols correspond to the  data:
  $ (\circ)\ tw=256s, \ (\bullet) t_w=353s, \ (\lozenge) t_w=4200s,
 \ (\vartriangleleft) t_w=6542s$.
The dashed lines are obtained from the best fits (see text for
details).
   In the inset
$C(t,t_w)-C(t_w,t_w)$  is plotted as a function of  time for
several $tw=250s;353s;503s;1120s;1624s;2583s;4200s$. The
correlation functions have been superposed by scaling $t-t_w$ by a
characteristic time $t_o(t_w)$ which is an increasing function of
$t_w$. }
    \label{effectivetemperature}
\end{figure}

Before discussing these experimental results we want to compare
them to the single frequency experiment performed on glycerol
\cite{Grigera}. In this experiment, $T_{eff}$ has been measured
only at $7Hz$. Thus we studied how $T_{eff}(7Hz,t_w)$ depends on
$t_w$ at $7Hz$ in our experiment. The time evolution of
$T_{eff}(7Hz,t_w)$ is plotted as a function of $t_w$ in
fig.\ref{effectivetemperature}a). The time evolution of
$T_{eff}(2Hz,t_w)$ is also plotted just to show the large
temperature difference between two frequencies. Let us consider
the evolution at $7Hz$ only.   As in the experiment of
ref.\cite{Grigera}, we confirm the fact that the violation is
observed even if $\omega t_w >> 1$, which is in contrast with
theoretical predictions. The biggest violation is for short times
after the quench where the effective temperature is surprisingly
huge: around 800K at 7Hz and $t_w=300s$. In the experiment on
glycerol the first data reported are for $t_w>1000s$. Thus if we
consider only data at $t_w>1000s$ in
fig.\ref{effectivetemperature}a) we see that our results are close
to those of ref.\cite{Grigera}. Indeed at $t_w=1000s$  we find in
our experiment $(T_{eff}-T_f)/(T_g-T_f)\simeq 2.4$ The glycerol
data give $(T_{eff}-T_f)/(T_g-T_f)\simeq 1$. Thus the relative
violations of FDT at $7Hz$ are very close in glycerol and
polycarbonate. However it would be interesting to check whether at
shorter times and at lower frequencies large  $T_{eff}$ could be
observed in glycerol too.

In order to compare with theoretical predictions
\cite{Kurchan,Peliti} and recent spin glass experiment
\cite{Herisson} we may plot the integrated response $R(t,t_w)$ as
a function of the correlation $C(t,t_w)$. The latter is obtained
inserting  measured $T_{eff}(f,t_w)$ in eq.\ref{SZ} and by Fourier
transforming this equation. $R(t,t_w)$ can be computed by Fourier
transforming $Real[Z(\omega,t_w)]$. FDR now takes the form
\cite{Peliti}:

\begin{equation}
-C(t,t_w)+C(t_w,t_w)= k_B \ T_{eff}(t,t_w) \ R(t,t_w)
\label{correlation}
\end{equation}

In the inset of  fig.\ref{effectivetemperature}b), we see that for
$t_w> 300s$ the shape of the decay of $C(t_w, t)$ remains
essentially the same. Indeed data for different $t_w$ can be
scaled onto a single master curve by plotting $C(t_w, t)$ as a
function $(t-t_w)/t_o(t_w)$, where  $t_o(t_w)$ is an increasing
function of $t_w$: approximately $t_o(t_w) \propto log(t_w)$ for
$t_w>500s$. The self-similarity of correlation functions, found on
our dielectric data, is a characteristic of the universal picture
of aging \cite{Mezard,Peliti,Kob1,Kob2,Berthier2}, which  has been
also observed in spin-glass experiment \cite{Herisson} and in the
structure function of the dynamic light scattering of colloidal
gels \cite{Weitz}. Thus our results confirm that this picture of
aging applies also to the polymer dielectric measurements. To
further investigate this aging,  we plot, in
fig.\ref{effectivetemperature}b), $R(t,t_w)$ as a function
$(-C(t,t_w)+C(t_w,t_w))/k_B$ at different $t_w$. The slope of this
graph gives $1/T_{eff}$. The symbols correspond to the data
whereas the dashed line are obtained by inserting  the best fit of
$T_{eff}$ in  eq.\ref{fitTeff}, in eq.\ref{SZ}. We clearly see
that  data at small $C(t,t_w)$ asymptotically converge to an
horizontal straight line, which means that the system has an
infinite temperature. At short time, large $C(t,t_w)$,  FDT
prediction is recovered (continuous straight line of slope
$1/T_f$). This result is quite different to what has been observed
in recent experiments on spin glasses where $T_{eff}\simeq 5 T_g$
has been measured\cite{Herisson}. In contrast infinite $T_{eff}$
has been observed during the sol gel transition \cite{Bellon} and
in numerical simulation of domain growth phenomena\cite{Barrat}.

\subsection{ Statistical analysis of the noise}

\begin{figure}[!ht]

 \centerline{\hspace{1cm} \bf (a) \hspace{8cm} (b) }

\begin{center}
        \includegraphics[width=8cm]{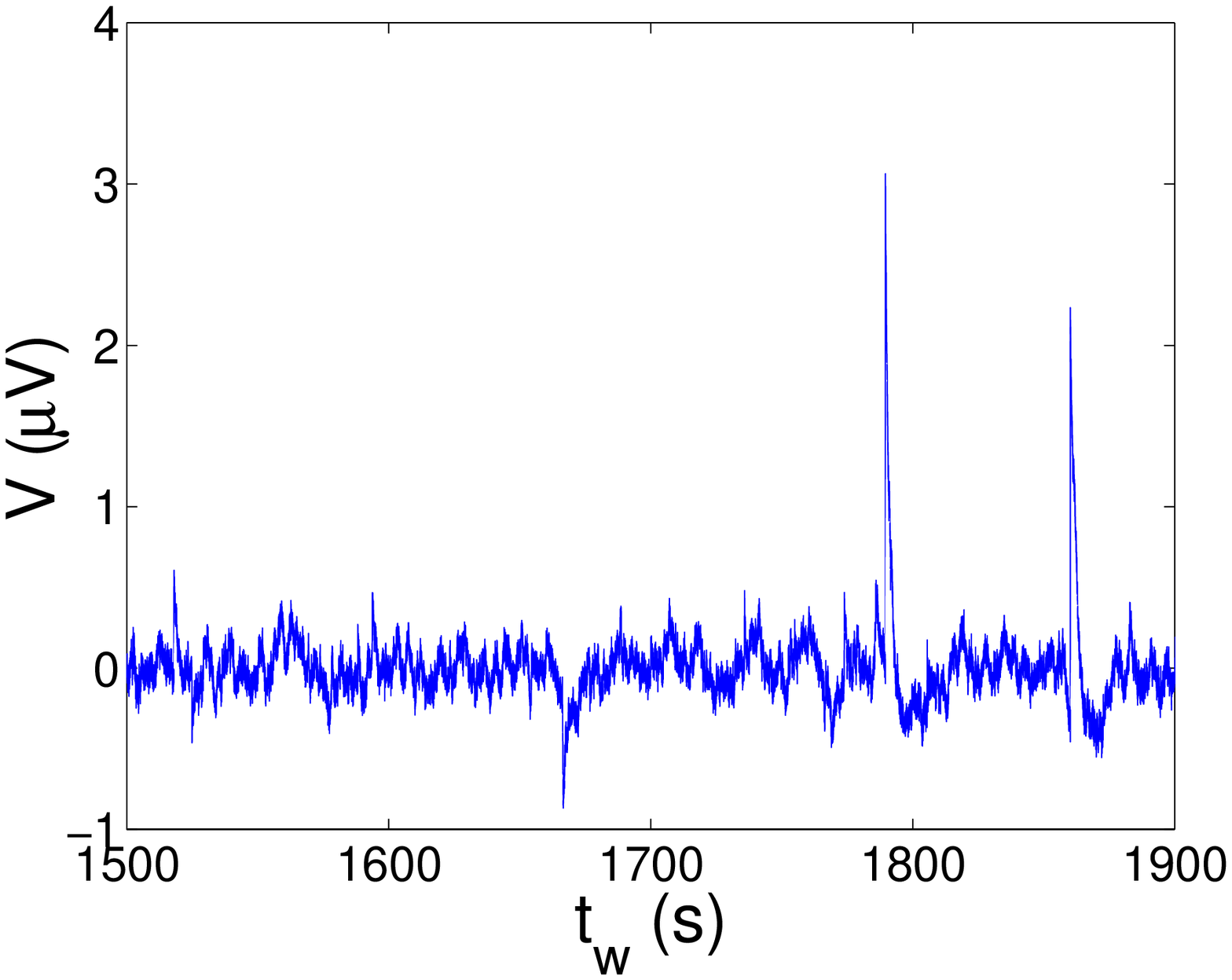}
        \hspace{1mm}
        \includegraphics[width=8cm]{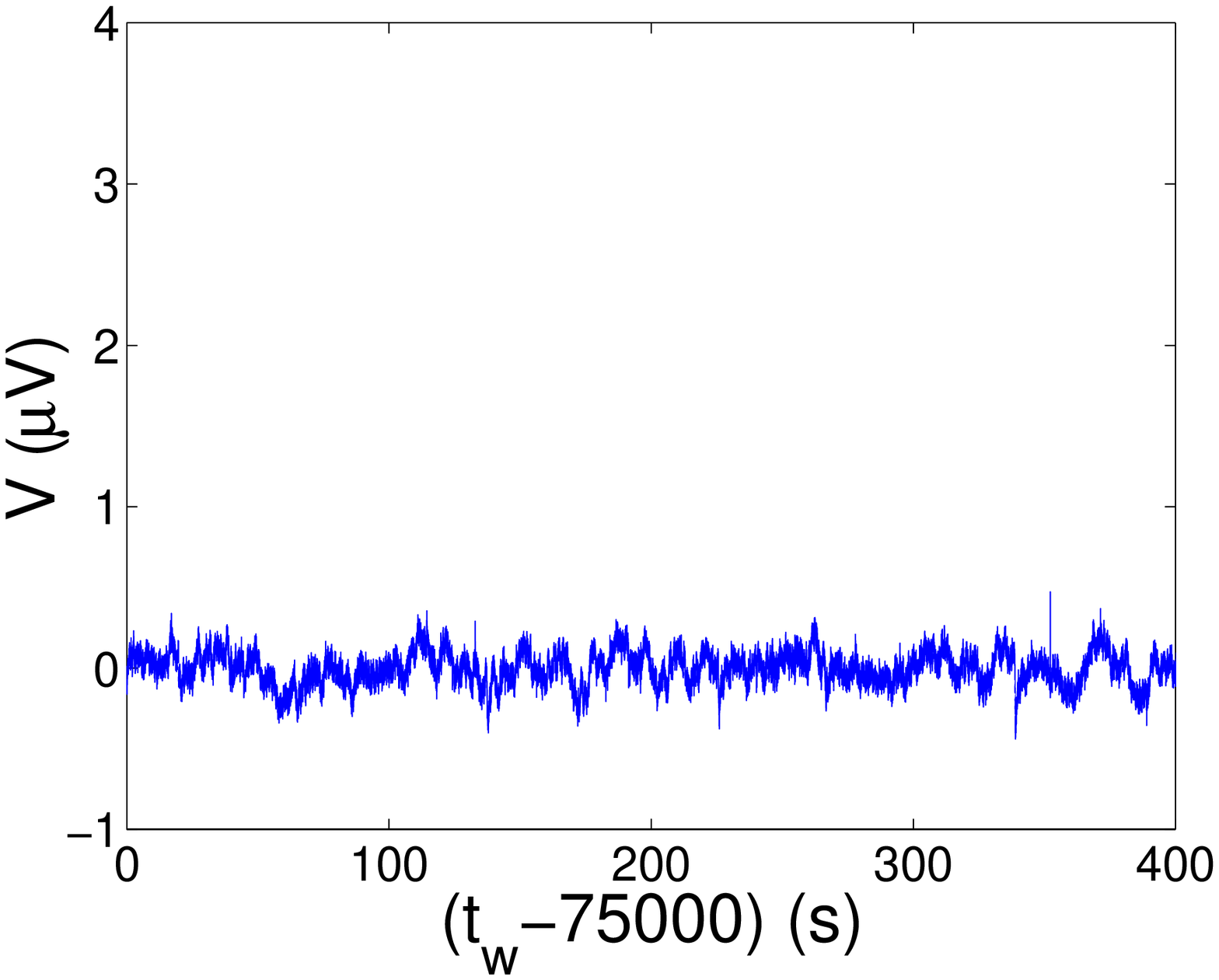}
    \end{center}

\caption{ {\bf Voltage noise signal in polycarbonate} Typical
noise signal of polycarbonate measured at $1500s<t_w<1900s$ (a)
and $t_w>75000s $ (b)}
 \label{signalpolyca}
\end{figure}

In order to understand the origin of such large deviations in our
experiment we have analyzed  the noise signal. We find that the
signal is characterized by  large intermittent events which
produce  low frequency spectra proportional to $f^{-\alpha}$ with
$\alpha \simeq 2$. Two  typical signals recorded at
$1500s<t_w<1900s$ and $t_w>75000s$  are plotted in
fig.\ref{signalpolyca}. We clearly see that in the signal recorded
at $1500s<t_w<1900s$  there are very large bursts which are on the
origin  of the frequency spectra discussed in the previous
section. In contrast in the signal (fig.\ref{signalpolyca}b),
which was recorded at $t_w>75000s$ when FDT is not violated, the
bursts are totally disappeared.

\begin{figure}[!ht]
\centerline{\epsfxsize=0.6\linewidth
\epsffile{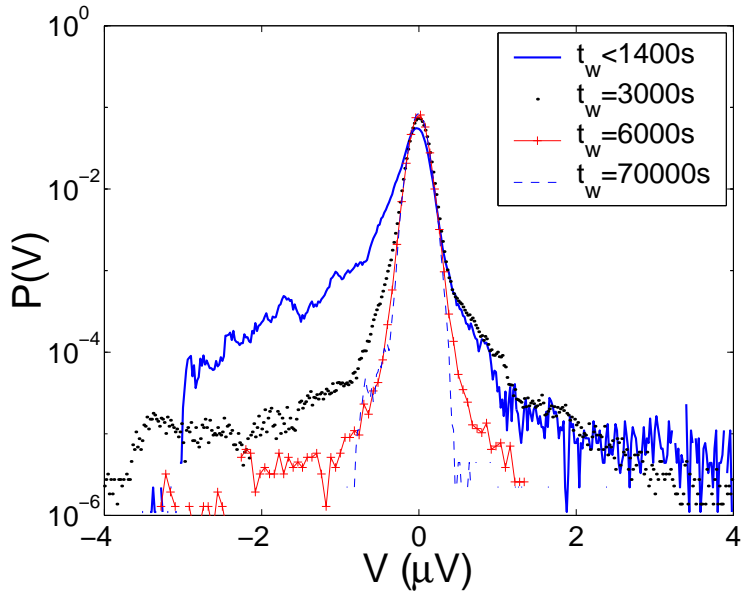}} \caption{ {\bf PDF of voltage
noise in polycarbonate} Typical PDF of the noise signal of
polycarbonate measured at various  $t_w$  }
 \label{PDFpolyca}
\end{figure}

As  for Laponite we have studied the PDF of the signal as a
function of $t_w$ for   polycarbonate. The results are shown in
fig.\ref{PDFpolyca}. We clearly see that the PDF, measured at
small $t_w$, has very high tails which becomes smaller and smaller
at large $t_w$. Finally the Gaussian profiles is recovered after
$24h$. This strongly intermittent dynamics is reminiscent of the
intermittence observed in the  local measurements of polymer
dielectric properties \cite{Israeloff_Nature} and in the slow
relaxation dynamics of a colloidal gel \cite{Mazoyer}

\section{Discussion and conclusions }

Let us resume the main results of the two experiments described in
the previous sections. We have seen that dielectric measurements
of  Laponite, during the sol-gel transition, and of polycarbonate,
 after a temperature quench, show a strong violation of FDT. The
effective temperature defined by eq.\ref{Teff} is huge at small
$t_w$ and slowly relaxes towards the bath temperature. In contrast
to theoretical predictions  the violation is observed even at
$\omega t_w \gg 1$  and it may last for more than $3h$ for
$f>1Hz$. We have  then  investigated the beahvior of the noise
signals and we have shown that the huge $T_{eff}$ is produced by
very large intermittent bursts which are at the origin of the low
frequency power law decay of noise spectra. Furthermore we have
also shown that for both materials the statistic of this event is
strongly non Gaussian when FDT is violated and slowly relaxes to a
Gaussian one at very long $t_w$. Thus these two very different
materials have a very similar relaxation dynamics, characterized
by a strong intermittency. This strongly intermittent dynamics is
reminiscent of the intermittence observed in the  local
measurements of polymer dielectric properties.
\cite{Israeloff_Nature}. Furthermore recent measurements done,
using  time resolved correlation in diffusing wave spectroscopy,
have shown a strong intermittency in the slow relaxation dynamics
of a colloidal gel \cite{Mazoyer}. This kind of beahvior can
indeed be interpreted on the basis of the trap model \cite{trap},
which predicts non trivial violation of FDT associated  to an
intermittent dynamics. The system evolves in deeper and deeper
valleys on the energy landscape. The dynamics is fundamentally
intermittent because either  nothing moves or there is a jump
between two traps. In our case these jumps could explain the
presence in the dielectric voltage noise of very large and rare
peaks with a slow relaxation after the jump.  Clear  answers to
this question can be given by a detailed study of the statistics
of the time intervals between large peaks. This work is in
progress.

This work clearly shows the importance of associating thermal
noise and response measurements. Many questions remain opened on
the subject of FDR in out of equilibrium systems. One may wonder
whether different couples of conjugated variables give the same
$T_{eff}$ as defined by eq.\ref{Teff}. For example FDR measured on
the rheological properties of Laponite show no violation of FDT
\cite{BellonD}. The reasons of these differences between
electrical and mechanical measurements are unclear and much work
is necessary to give new insight on these problems.

{\noindent \bf Acknowledgments }

We acknowledge useful discussion with J. Kurchan and  J.P.
Bouchaud. We thank P. Metz and F. Vittoz for technical support.
This work has been partially supported by the R\'egion
Rh\^one-Alpes contract ``Programme Th\'ematique : Vieillissement
des mat\'eriaux amorphes''.

\end{document}